\PassOptionsToPackage{table,x11names}{xcolor}
\documentclass[]{article}
\usepackage{dcase2022_techrep,amsmath,graphicx,url,times,booktabs, tabularx}
\usepackage{xcolor}
\usepackage{tikz}

\title{DCASE 2022: COMPARATIVE ANALYSIS OF CNNs FOR ACOUSTIC SCENE CLASSIFICATION UNDER LOW-COMPLEXITY CONSIDERATIONS}

\name{Josep Zaragoza-Paredes$^{1}$,
       Javier Naranjo-Alcazar$^{2}$,
       Valery Naranjo$^{1}$, 
       Pedro Zuccarello$^{2}$
       }
\address{$^1$ Universitat Politecnica de Valencia, ETSIT, Valencia, Spain, \{jozapa@teleco, vnaranjo@dcom\}.upv.es \\          
$^2$ Instituto Tecnologico de Informatica, Valencia, Spain 
\{jnaranjo, pzuccarello\}@iti.es\\ 
  }

\begin{document}
\def\arraystretch{1.5}
\setlength\arrayrulewidth{1pt}

\ninept
\maketitle

\begin{sloppy}

\begin{abstract}
Acoustic scene classification is an automatic listening problem that aims to assign an audio recording to a pre-defined scene based on its audio data. Over the years (and in past editions of the DCASE) this problem has often been solved with techniques known as ensembles (use of several machine learning models to combine their predictions in the inference phase). While these solutions can show performance in terms of accuracy, they can be very expensive in terms of computational capacity, making it impossible to deploy them in IoT devices. Due to the drift in this field of study, this task has two limitations in terms of model complexity. It should be noted that there is also the added complexity of mismatching devices (the audios provided are recorded by different sources of information). This technical report makes a comparative study of two different network architectures: conventional CNN and Conv-mixer. Although both networks exceed the baseline required by the competition, the conventional CNN shows a higher performance, exceeding the baseline by 8 percentage points. Solutions based on Conv-mixer architectures show worse performance although they are much lighter solutions.
\end{abstract}


\section{Introduction}
\label{sec:intro}

Acoustic Scene Classification is understood as the task within the field of machine listening that aims at extracting context from audio data. This context is achieved by assigning a pre-defined scene to an audio clip \cite{naranjo2020acoustic, perez2019cnn, battaglino2016acoustic, singh2022passive, pham2022wider}. Thus, the audio is referenced to the location where it is being recorded (park, airport, shopping mall, among others). Due to the characteristics of this problem, it can be defined as a supervised classification problem.

Since the first edition of the DCASE Challenge in 2013, this problem has always appeared as Task 1 \cite{Giannoulis2013, Mesaros2016_EUSIPCO, DCASE2017challenge}. However, it has not always been presented in the same manner as it has been modified over the years to reflect different constraints when deploying machine listening solutions, two of them being the appearance of the open set \cite{Zhu2019} or the limitation of mismatch devices \cite{Mesaros2018_DCASE}. Over the years, one of the most common practices by participants was the use of ensembles \cite{Huang2019, Ding2019, Haocong2019}. The idea behind this solution lies in the use of various machine learning models to improve the performance of the final system. By combining the output of several systems, a more robust prediction is obtained \cite{perez2019cnn}. However, these solutions are limited by the number of operations that need to be performed to obtain a final output. This makes it impossible (or challenging) to deploy these solutions in IoT devices. 

This year's edition (2022) presents two considerations for the system to be taken as valid:

\begin{itemize}
    \item The maximum number of parameters is set to 128K and the used variable type is fixed into INT8
    \item The maximum number of multiply-accumulate operations is set to 30M
\end{itemize}

With this in considerations, it has been decided to design two solutions:

\begin{itemize}
    \item Solution based on a convolutional module composed of a conventional convolutional layer and a separate convolutional layer (Conv-Sep)
    \item Solution based on the module known as the Conv-mixer
\end{itemize}

Both solutions have the same structure (number of filters, pool sizes, dropouts, etc.) so that they can be fully comparable (for more information see~\ref{subsec:system}). To meet the requirements of the task, the models have been converted to a TFLite format with 8-bit quantization. 

In terms of the performance of the proposed systems, both solutions exceed the baseline proposed by the organisation. While the Conv-Sep solution shows better performance with the same number of filters, it also shows a higher number of MMACs and parameters. All these results are discussed in Section~\ref{sec:results}.

The rest of the technical report is divided as follows: section~\ref{sec:method} presents the different blocks of the system (audio representation, machine learning model and training process). Section~\ref{sec:results} shows the results. Finally, section~\ref{sec:conclusion} concludes our work.

\section{Method}
\label{sec:method}

The following subsections explain the modules that make up the system presented to the Challenge.

\subsection{Audio representation}\label{subsec:audio}

The first module of the system consists of obtaining a time-frequency representation of the audio, $F \times T$, where $F$ is the number of frequency bins and $T$ is the time frames. For temporal analysis, a window of 40ms with 50\% overlap is used. The sample rate is set to 44100 Hz. For frequency resolution, it has been decided to use a bank of 64 Mel filters. Finally, the logarithm is performed to obtain the log-Mel spectrogram \cite{Kim2021b, Yang2021}. This configuration crates a $64 \times 51$ audio log-Mel spectrogram from an audio of 1 second duration. This  module has been implemented using the Librosa Python package.

\subsection{System description}\label{subsec:system}

\begin{table}[]
\centering
\begin{tabular}{c}
\hline
\textbf{Model architecture}     \\ \hline
Log-Mel spectrogram representation $64 \times 51 \times 1$ \\ \hline
Convolutional module (Number of filters)  \\ \hline
MaxPooling(1,4)        \\ \hline
Dropout(0.3)           \\ \hline
Convolutional module (Number of filters)  \\ \hline
MaxPooling(1,2)        \\ \hline
Dropout(0.3)           \\ \hline
Global Average Pooling \\ \hline
Dense(10)              \\ \hline
\end{tabular}
\caption{Base architecture of the model used in the Challenge. The Convolutional module corresponds to the module explained in the different sections of subsection~\ref{subsec:system}. The different values in parenthesis correspond to the hyperparameters used. In the MaxPooling layer the pool size is defined, in the Dropout layer the rate is specified and in the Convolutional module the number of filters. The Dense layer has 10 units, corresponding to the number of classes to predict}
\label{tab:model}
\end{table}

The model used has a VGG-style architecture \cite{simonyan2014very}. Convolutional layers are interspersed with pooling and dropout layers. In this paper, two convolutional layer structures are compared. The first one is based on a two-layer convolutional module and the second one is based on the Conv-mixer module. The base model architecture can be seen in Table~\ref{tab:model}.

\subsubsection{Conv-Sep model}\label{subsec:conv-sep}

This architecture is based on the VGG architecture. Two convolutional layers are stacked followed by pooling layers (dimensionality reduction) and dropout layers (avoiding overfitting). However, a number of modifications have been made with respect to a conventional VGG network. ReLU activation layers are replaced by ELU layers. To meet the complexity requirements imposed in the Challenge, the second convolutional layer is replaced by a Separable convolution layer. Thus, the convolutional module can be seen in Table~\ref{tab:conv_sep}

\begin{table}[]
\centering
\begin{tabular}{c}
\hline
\textbf{Convolutional module Conv-Sep}              \\ \hline
Conv Layer (number of filters)           \\ \hline
Batch Normalization                      \\ \hline
ELU                                      \\ \hline
Separable Conv Layer (number of filters) \\ \hline
Batch Normalization                      \\ \hline
ELU                                      \\ \hline
\end{tabular}
\caption{Architecture of the convolutional module of the Conv-Sep model. The number of filters in the two convolutional layers is the same}
\label{tab:conv_sep}
\end{table}


\subsubsection{Conv-mixer model}\label{subsec:conv-mixer}

The Conv-mixer module was presented in \cite{trockman2022patches}. It consists of a module with two convolutional layers, the first one being DepthWise and the second one PointWise. The two layers are followed by a GELU activation layer and a BatchNormalization layer. In addition, there is a residual connection between the previous point of the first layer and the result obtained from this layer. Prior to the first convolutional module, path embedding is performed. The path size is set to \textcolor{blue}{1}. For more details, please see \cite{trockman2022patches}.

\subsection{Experimental details}\label{subsec:exp_details}

\subsubsection{Training details}\label{subsubsec:training}

The following configuration has been used in the training phase:

\begin{itemize}
    \item The maximum number of epochs is set to 500
    \item The training is stopped if accuracy is not improved within 30 epochs
    \item The learning rate is reduced by a factor of 0.5 if the accuracy is not improved within 15 epochs
    \item The optimizer used is Adam
    \item Loss function is set to categorical crossentropy
\end{itemize}

All models have been implemented with the ML library Tensorflow.

\subsubsection{Dataset}\label{subsubsec:dataset}

The dataset used in this task is the TAU Urban Acoustic Scenes 2022 Mobile, development dataset. The dataset is composed of recordings from 4 different real devices and 11 simulated devices. The recordings have been made in 12 cities, 10 of which are present in the development set. The audios have a duration of 1 second. The training set is composed of 40 hours of audio provided in mono, 44100Hz sampling rate and 24-bit format. The classes present are 10 (as in the other editions). In summary, the training set consists of 139970 segments and the test set consists of 29680 segments. 

\subsection{Model quantization}\label{subsec:quantization}

Following the requirements of the task. The weights of the model have undergone a Post-training quantization. For this procedure, the tools present in Tensorflow have been used to convert and quantise models to the TFLite format. 

\begin{table*}[]
\centering
\begin{tabular}{cccccc}
\hline
Model              & Number of filters & Accuraccy(\%)& Log Loss & MMACs & Number of parameters\\ \hline \hline
\rowcolor{Blue3!20!}Challenge baseline & [16-32]                  &      42.90     &  1.58   & 	29,234,920 & 46,512 \\ \hline
\rowcolor{SeaGreen3!25!} Conv-Sep                & [40-40]                  &    50.00     &   \textbf{1.44}  & \textbf{20,306,320} & \textbf{20,088}\\ \hline
Conv-mixer     & [40-40]            &   46.32   &  1.50    & 17,979,280  & 10,515 \\ \hline
\rowcolor{SeaGreen3!25!} Conv-Sep & [48-48] & \textbf{50.57} & \textbf{1.44} & 28,570,080 & 28,320 \\ \hline
Conv-mixer* & [48-48] & 47.26 & 1.64 & 24,671,712 & 14,139 \\ \hline
\rowcolor{SeaGreen3!25!} Conv-Sep & [32-64] & 50.09  & 1.45 & 23,138,944 & 26,544 \\ \hline
Conv-mixer & [32-64] & 46.76 & 1.64 & 15,895,168  & 12,683  \\ \hline
\rowcolor{Red3!20!} Conv-Sep & [64-64] & 51.12  & 1.49 & 49,300,096 & 49,008 \\ \hline
Conv-mixer & [64-64] & 47.79 & 1.70 & 41,153,152  &  22,923 \\ \hline
\hline
\end{tabular}
\caption{Results obtained using proposed network with convolutional layer configuration and conv-mixer before quantization. The model indicated as * was the one submitted to the Challenge in TFLite format and int8 quantization}
\label{tab:results}
\end{table*}

\section{Results}
\label{sec:results}

The results obtained can be seen in Table~\ref{tab:results}. The models that have been submitted to the Challenge are shown in green. 

An analysis of each filter configuration has been carried out for each of the models. As can be seen, the number of parameters and MMACs is drastically lower with the Conv-mixer configuration. However, the most promising results are obtained for the model with the Conv-Sep module.

As can be seen, the configuration of two Conv-Sep modules with 48 filters per module. The accuracy improves the baseline of the Challenge with about 8 perceptual points and the Log Loss is reduced with a value of 0.14.

Solutions based on Conv-mixer modules show a worse performance compared to Conv-Sep. The Conv-mixer based solution presents a great improvement in terms of complexity, being these solutions very lightweight. However, the best solution only improves by 5\% the system. This configuration sometimes improves the accuracy but worsens the Log Loss.

\section{Conclusion}
\label{sec:conclusion}

The results obtained in this submission show that the performance of an ASC system can be improved while reducing its complexity. Two architectures based on convolutional layers are analysed in this paper. The Conv-mixer based solutions show an improvement in complexity. However, no configuration has been found to improve the Conv-Sep setup. Solutions based on VGG architectures (interleaving convolutional and separable layers) show a great performance-complexity trade-off.


\section{ACKNOWLEDGMENT}
\label{sec:ack}

\bibliographystyle{IEEEtran}
\bibliography{refs}
%
%
%
%
%
%
%
%
%

\end{sloppy}
\end{document}